\documentclass[doublespacing]{elsart}

\usepackage{graphicx}
\usepackage{amssymb}

\begin{document}
\begin{frontmatter}


\journal{SCES'2001: Version 1}


\title{Optical spectroscopy of $\rm \bf (La,Ca)_{14}Cu_{24}O_{41}$
spin ladders: comparison of experiment and theory}

%
%
%
%
%
%

\author[koeln_ex]{M.\ Gr\"{u}ninger\corauthref{1}}
\author[koeln_ex]{M.\ Windt}
\author[augsburg]{T.\ Nunner}
\author[koeln_th]{C.\ Knetter}
\author[koeln_th]{K.P.\ Schmidt}
\author[koeln_th]{G.S.\ Uhrig}
\author[augsburg]{T.\ Kopp}
\author[koeln_ex]{A.\ Freimuth}
\author[koeln_ex,paris]{U.\ Ammerahl}
\author[koeln_ex,aachen]{B.\ B\"{u}chner}
\author[paris]{A.\ Revcolevschi}

%

\address[koeln_ex]{II. Physikalisches Institut, Universit\"{a}t zu K\"{o}ln, 50937 K\"{o}ln, Germany}
\address[augsburg]{Experimentalphysik VI, Universit\"{a}t Augsburg, 86135 Augsburg, Germany}
\address[koeln_th]{Institut f\"{u}r Theoretische Physik, Universit\"{a}t zu K\"{o}ln, 50937 K\"{o}ln, Germany}
\address[aachen]{II. Physikalisches Institut, RWTH-Aachen, 52056 Aachen, Germany}
\address[paris]{Laboratoire de Physico-chimie, Universit\'e Paris-Sud, 91405 Orsay, France}

%
%
%
%


%
%
%
%

\corauth[1]{Email of Corresponding Author: grueninger@ph2.uni-koeln.de}


\begin{abstract}
Transmission and reflectivity of $\rm
La_{x}Ca_{14-x}Cu_{24}O_{41}$ two-leg spin-1/2 ladders were
measured in the mid-infrared regime between 500 and
12\,000\,cm$^{-1}$. This allows us to determine the optical
conductivity $\sigma_1$ directly and with high sensitivity. Here
we show data for $x$=4 and 5 with the electrical field polarized
parallel to the rungs ($E \, || \, a$) and to the legs ($E \, ||
\, c$). Three characteristic peaks are identified as magnetic
excitations by comparison with two different theoretical
calculations.
\end{abstract}

%
%



\end{frontmatter}

%
%
%
%
%

The quantum nature of magnetic excitations in spin-1/2 systems
and in particular the role of quantum fluctuations in low
dimensions are a fascinating subject. Antiferromagnetic (AF)
S=1/2 Heisenberg ladders represent an intermediate step between
one-dimensional (1D) chains and the 2D CuO$_2$ layers of undoped
high-$T_c$ superconductors. The elementary excitations of the
ladders can be described as triplets or as interacting spinons.
Topics of current interest are theoretical predictions of
2-triplet bound states \cite{sushkovtrebstbrenig}, the size of
the exchange coupling along the rungs ($J_\perp$) and the legs
($J_\parallel$) as well as the role of the ring exchange $J_{\rm
cyc}$ \cite{johnston}. We address these issues in $\rm
La_xCa_{14-x}Cu_{24}O_{41}$ which contains layers with
Cu$_2$O$_3$ two-leg AF S=1/2 ladders \cite{windt}.

\begin{figure}
\centering
\includegraphics[width=5.8cm,clip]{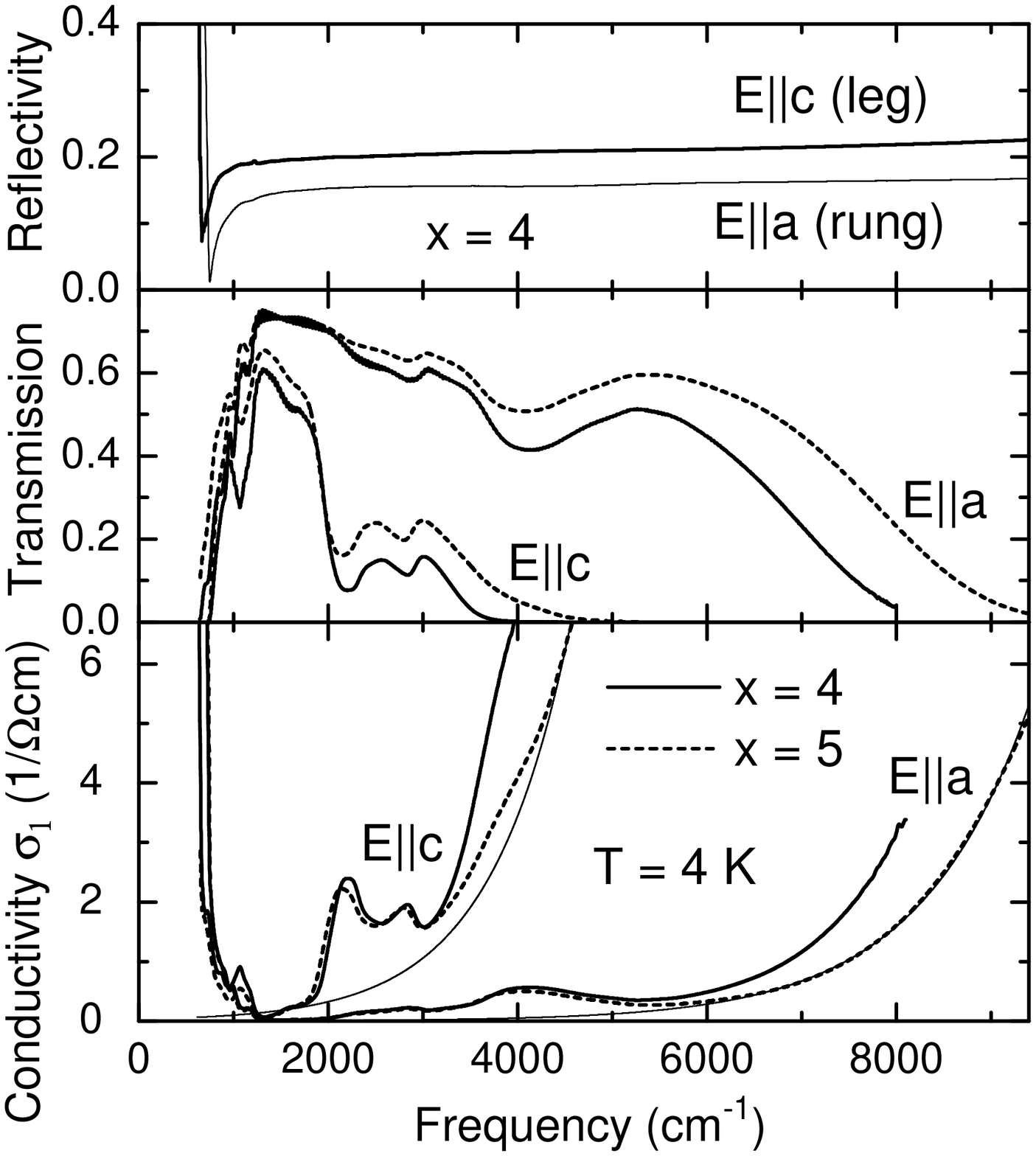}
\caption{Reflectivity, transmission and optical conductivity of
$\rm La_xCa_{14-x}Cu_{24}O_{41}$. Solid lines: $x$=4; dashed:
$x$=5; thin lines in lower panel: exponential fits to the
electronic background. Transmission sample thicknesses:
$\approx$60\,$\mu$m ($\approx$44\,$\mu$m) for $x$=4 (5). }
\label{fig1}
\end{figure}

A La content of $x$=6 corresponds to nominally undoped samples,
i.e.\ Cu$^{2+}$, but single phase crystals were obtained only for
$x \lesssim$ 5 \cite{udo}. Reflectivity and transmission data for
$x$=5 and 4 at 4\,K are plotted in Fig.\,\ref{fig1} along with the
deduced real part $\sigma_1$ of the optical conductivity. Except
for the strong phonon signature at low frequencies the
reflectivity is featureless, demonstrating that reflectivity
measurements with subsequent Kramers--Kronig transformation are
not adequate to resolve small values of $\sigma_1$. The
transmission, however, is much more sensitive to weak absorption
and combining transmission and reflectivity one can determine
$\sigma_1$ most accurately.

\begin{figure}
\centering
\includegraphics[width=5.8cm,clip]{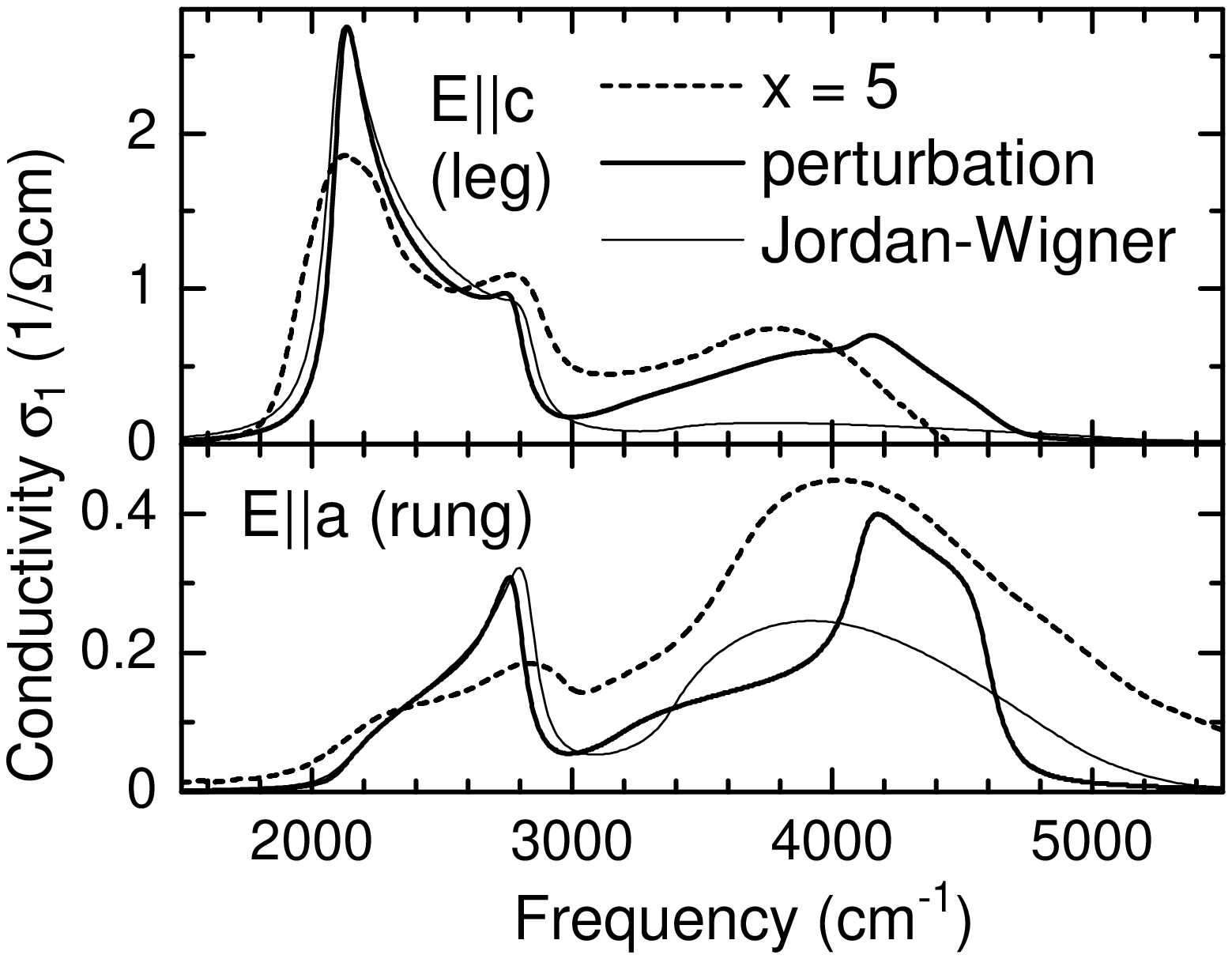}
\caption{Magnetic contribution to $\sigma_1$ of $\rm
La_{5}Ca_{9}Cu_{24}O_{41}$ (dashed lines) compared with
calculations using optimized perturbation (thick lines, $J_\perp
= J_\parallel = 1020$\,cm$^{-1}$) and Jordan-Wigner fermions (thin
lines, 1100\,cm$^{-1}$), respectively. The assumed phonon energy
is 600\,cm$^{-1}$. An exponential electronic background was
subtracted from the measured data of Fig.\,1. Note the small
values of $\sigma_1
\leq 3 \, (\Omega$cm$)^{-1}$.}
\label{fig2}
\end{figure}

The spectra can be divided into 3 different regimes. Below
$\approx$1300\,cm$^{-1}$ the rise of $\sigma_1$ is due to phonon
absorption. The high frequency behavior is dominated by an
electronic background that increases with hole doping, i.e.\
decreasing $x$. To analyze the peaks in the intermediate region
we subtracted this background using an exponential fit (thin
lines in Fig.\,\ref{fig1}). After subtraction the remaining
features are almost independent of $x$ (see Fig.\,\ref{fig2} for
$x$=5). We interpret these excitations in terms of phonon-assisted
two-magnon absorption \cite{lorenzana95} which has been used to
describe $\sigma_1$ of the undoped 2D cuprates (e.g.\ $\rm
YBa_2Cu_3O_6$ \cite{gruen}) and of the 1D S=1/2 chain $\rm
Sr_2CuO_3$ \cite{lorenzana97}. Due to spin conservation two
magnons are excited. The simultaneous excitation of a phonon
provides the symmetry breaking necessary to bypass the selection
rule and it guarantees momentum conservation
\cite{windt,lorenzana95,gruen}. Since the exchange coupling in the
chains is $\approx 2$ orders of magnitude smaller than in the
ladders, we attribute the observed absorption to the ladders.


In Fig.\,\ref{fig2} we compare the magnetic contribution to
$\sigma_1$ of the lowest nominal doping $x$=5 (dashed lines; one
hole per formula unit) with 2 different theoretical calculations.
One approach is related to 1D spinon physics and describes the
spins in terms of Jordan--Wigner fermions with a long-ranged
phase factor (thin lines in Fig.\,\ref{fig2}). The other approach
starts from isolated singlets on each rung, i.e.\
$J_\parallel$=0, with local triplet excitations. Using continuous
unitary transformations \cite{knetter}, finite $J_\parallel$ is
then treated as a perturbation that creates delocalized, dressed
triplets (thick lines in Fig.\ \ref{fig2}). Concerning the
dispersion of the elementary excitation (triplet or ``magnon''),
the differences between both theories are $\lesssim$10--20\%
\cite{windt}. Both show a dispersing two-triplet bound state with
$S_{\rm tot}$=0 that leaves the two-triplet continuum at
$k$$\gtrsim$0.3$\pi$. The maximum of this bound state at
$k$$\approx$$\pi/2$ and its minimum at $k$=$\pi$ yield van-Hove
singularities in the density of states that cause the 2 peaks at
2800 and $\rm 2140\,cm^{-1}$, respectively. Both theories are in
excellent agreement with the experimental data for $J_\parallel /
J_\perp
\!\approx \!1$--1.2 with $J_\parallel \!\approx\!
1020$--1100\,cm$^{-1}$. Further confirmation of our interpretation
is the reduced spectral weight of the peak at 2140\,cm$^{-1}$ for
$E \, || \, a$ caused by a selection rule arising from symmetry
\cite{windt}. We have thus verified the theoretical predictions
of a two-triplet bound state \cite{sushkovtrebstbrenig}. Finally,
the broad peak at around 4000\,cm$^{-1}$ is identified with the
two-triplet continuum.

A ratio of $J_\parallel / J_\perp$$\approx$1 seems to be in
conflict with several former results of other techniques,
proposing $J_\parallel / J_\perp \gtrsim 1.5$ (see discussion in
\cite{johnston}). Such large values can be excluded on the basis
of our results \cite{windt}. The introduction of a ring exchange
$J_{\rm cyc}\approx 0.15 \, J_\parallel$ resolves this issue in
favor of $J_\parallel / J_\perp$$\approx$1--1.1 \cite{matsuda}.

In conclusion, the existence of a two-triplet bound state is
verified in the two-leg S=1/2 ladders of $\rm
La_{x}Ca_{14-x}Cu_{24}O_{41}$ ($x$=5 and 4). We obtain the values
of the exchange constants
$J_\parallel$$\approx$1020--1100~cm$^{-1}$ and $J_\parallel /
J_\perp$$\approx$1--1.2.

This project is supported by the DFG (FR 754/2-1, SP 1073 and SFB
484), by the BMBF (13N6918/1) and by the DAAD within the scope of
PROCOPE.

%
%
%
%

%
%
%
%

\end{document}